\documentstyle[prb,epsf,aps]{revtex}
\begin{document}

\preprint 
\draft

\title{Virial Expansions, Exclusion Statistics, and the Fractional
Quantum Hall Effect}
\author{Kahren Tevosyan and A. H. MacDonald}
\address{Department of Physics, Indiana University, Bloomington
IN 47405, USA}

\date{\today}
\maketitle

{\tightenlines
\begin{abstract}

We report on a study of virial expansions for interacting electrons in the 
lowest Landau level of a two-dimensional electron gas.   For 
hard-core-model interactions, we derive 
analytic results valid at low temperatures and 
filling factors smaller than $1/3$ and comment on their relationship
with virial expansions for exclusion statistics models.  
In the high temperature limit the virial coefficients reduce
to those for a non-interacting electron system.  For the 
first five virial coefficients, the crossover 
between low and high temperature limits has been studied 
numerically by using partition functions obtained from small system 
exact diagonalization calculations.  Our results show that the 
exclusion statistics description of fractional Hall thermodynamics
breaks down when the temperature exceeds a small fraction of 
the gap temperature.

\end{abstract}
}

\section{Introduction}

The spectrum of the kinetic energy operator for a non-interacting 
electron in two dimensions in a perpendicular magnetic field
consists of Landau levels with 
degeneracy and energy separation both proportional to the 
magnetic field strength $B$.  In strong magnetic fields, 
the situation where all electrons can be accommodated within 
the lowest Landau level can be realized experimentally
and affords an interesting archetype for strongly interacting 
electron systems.  In such systems, the many-particle kinetic
energy operator has a ground state degeneracy that grows exponentially
with system size, frustrating attempts to describe the 
system by treating electron-electron interactions perturbatively.  
The physical properties of two-dimensional electron systems (2DES) 
have a corresponding richness, exhibiting a number of unusual 
properties including the fractional quantum Hall\cite{fqhebooks}
effect (FQHE).  The FQHE is related to discontinuities in the 
chemical potential\cite{leshouches} of a 2DES at zero temperature.  Progress 
has been made in its theoretical understanding by a variety 
of approaches, including the use of variational
 many-electron wave functions,\cite{laughlin}
the idenfication of model Hamiltonians\cite{haldane,trugkiv}
for which the ground state
may be determined analytically, and the introduction of artificial
Chern-Simons magnetic flux quanta attached to each particle.
\cite{jaincf,hlr,kivelson,lopezfradkin}  
In this article we discuss an initial attempt to address finite-temperature 
properties of 2DES's in the FQHE regime using virial expansions.

Throughout this article we will restrict our attention to
the situation where the energy separation between Landau levels 
($\hbar \omega_c = \hbar e B / m^{*} c$) is much larger than
the thermal energy $k_B T$ and assume that the electronic system is 
completely spin polarized so that we can truncate the Hilbert
space to the lowest spin-polarized Landau level.   In Section II we introduce 
appropriate definitions for cluster and virial expansion coefficients 
in this FQHE regime.  Here we discuss the non-interacting electron
limit and some exact properties of the virial expansion that follow from 
particle-hole symmetry within the lowest Landau level.  For interacting
electron systems, the non-interacting results for the virial coefficients 
are recovered in the high-temperature limit where $k_B T$ is much larger 
than the typical interaction energy per particle.  
All interaction-dependent results discussed in this article 
are for the case of hard-core-model
electron-electron interactions.  This model can be regarded as 
the ideal model of the fractional quantum Hall effect since the  
seminal variational wave functions of Laughlin are
 exact in this case.\cite{haldane,trugkiv}
In Section III we derive exact results for the zero-temperature limit
of the virial coefficients of the hard-core model.  
As we discuss there, the thermodynamics in this regime corresponds precisely to 
that of an exclusion statistics gas.\cite{exclusionstats}
In order to study how the virial coefficients interpolate between
their high-temperature and low-temperature limits, we have 
obtained numerically exact results for the temperature dependence 
of the five leading virial coefficients by calculating 
finite-size partition functions for small numbers of electrons.
These results are presented and 
discussed in Section IV.  In Section V we discuss the degree to 
which fractional quantum Hall behavior is reflected 
in the virial coefficients we have evaluated.
Finally in Section VI we briefly summarize our results. 

\section{Virial Expansions in the Lowest Landau Level}

The virial expansion\cite{huang} can be derived from the 
expansion of the grand potential in powers of the fugacity,
$z \equiv \exp (\mu / k_B T)$.  This expansion is motivated 
in part by the fact that only its leading order term is non-zero
for the classical non-interacting gas.  Higher order terms in
the expansion are due to quantum effects and interactions.  
The grand potential is 
extensive and for the two-dimensional case, the dimensionless 
coefficients of the virial expansion are conventionally defined by 
expressing the grand potential per unit area in units of $ - k_B T 
/ \lambda^2$ where $\lambda = (2 \pi \hbar^2/ m^{*} k_B T)^{1/2}$ is
the thermal de Broglie wavelength and $m^{*}$ is the effective 
electronic mass.  When the Hilbert space for a quantum
system is truncated to the lowest Landau level, $m^{*}$ 
plays no role and a different normalization is more convenient.
We develop our virial expansion for the FQHE regime by expanding the 
grand potential as follows:
\begin{equation}
\Omega(T,z)\equiv -k_BT\ln Z=-k_B T N_\phi \sum_{l=1}^\infty b_l(T) z^l
\label{eq:cluster}
\end{equation}
Here $N_{\phi} = A B / \Phi_0$ is the degeneracy of the Landau level,
$A$ is the area of the system and $\Phi_0 = h c /e $ is the electron magnetic 
flux quantum.  The dimensionless coefficients $b_l(T)$ can be determined
by expanding $\Omega(T,z)$ when this function is known analytically,
by using perturbative cluster expansion approaches,
or iteratively from canonical ensemble 
partition functions evaluated with successively larger 
finite number of electrons.  We will refer to the coefficients 
$b_l(T)$ as cluster expansion coefficients.  
It follows from Eq.\ (\ref{eq:cluster}) that 
\begin{equation}
\nu (T,z)=\sum_{l=1}^\infty l b_l(T) z^l
\label{eq:filling}
\end{equation}
where $\nu (T,z) \equiv N(T,z)/N_{\phi}$ is the Landau level filling
factor at finite temperature.  The virial expansion is obtained from 
these two series by inverting Eq.\ (\ref{eq:filling}) to obtain 
$z$ as a power series in $\nu$ and inserting the result in 
Eq.\ (\ref{eq:cluster}).   The procedure is readily carried out to
any finite order with the result that 
\begin{equation}
\frac{-\Omega}{k_BT N_{\phi}}=\sum_{l=1}^{\infty} a_l(T) \nu^{l}
\label{eq:virialexp}
\end{equation}
where $a_l(T)$ is $l^{\rm th}$ virial coefficient.  

For 2DES's in the fractional quantum Hall regime a number 
of other thermodynamic functions are of interest.
We will comment extensively on the thermodynamic density of states:
\begin{eqnarray}
g_T &\equiv& \frac{d N}{d \mu} =
N_{\phi} \frac{z}{k_B T}\frac{ d \nu}{ d z} \nonumber\\\
&=&\frac{N_{\phi}}{k_BT} \sum_{l=1}^{\infty} g_l(T) \nu^ l ; 
\label{eq:tdos}
\end{eqnarray}
the free energy
\begin{equation}
F = \Omega  +  N k_B T \ln z = N_{\phi} k_B T \big[ \nu \ln (\nu) + \sum_l
f_l(T) \nu^l \big];
\label{eq:freeen}
\end{equation}
the entropy 
\begin{equation}
S = \frac{ - \partial F}{\partial T} = -
N_{\phi} k_B [\nu \ln (\nu) + 
\sum_{l=1}^{\infty} \frac{d(T f_l(T))}{dT} \nu^l];
\label{eq:entropy}
\end{equation}
and the heat capacity 
\begin{equation}
C = T \frac{ \partial S} { \partial T} 
 = - N_{\phi} k_B \sum_{l=1}^{\infty} T 
 \frac{d^2( T f_l(T))}{d T^2} \nu^l. 
\label{eq:heatcapcity}
\end{equation}
Expressions for $a_l(T)$, $g_l(T)$, and $f_l(T)$ in terms 
of the cluster expansion parameters $b_l(T)$ are listed in Table I.

It is instructive to compute the coefficients of these expansions
for the case of a non-interacting Fermi gas.  We choose the 
zero of energy at the kinetic energy of particles in the lowest
Landau level for convenience.  Then the grand potential is 
given by
\begin{equation}
\Omega_{NI} = - N_{\phi} k_B T \ln (1 + z).
\label{eq:omegani}
\end{equation}
Differentiating $\Omega_{NI}$ with respect to $\mu$ we recover the 
Fermi distribution function expression for the Landau level
filling factor:
\begin{equation}
\nu = z / (1 +z ).
\label{eq:fillni}
\end{equation}
Expanding either Eq.\ (\ref{eq:omegani}) or Eq.\ (\ref{eq:fillni})
we identify the cluster expansion coefficients:
\begin{equation}
b_l^{NI} = (-1)^{l+1}/l.
\label{eq:blni}
\end{equation}
Eq.\ (\ref{eq:blni}) should be compared with the result for cluster 
expansion coefficients of quantum non-interacting Fermi gases
in D space dimensions in zero magnetic
field;\cite{huang} $ b_l \propto (-1)^{l+1} l^{-1-D/2}$.
As expected, in the strong field limit, Landau level degeneracy 
transforms thermodynamic properties of two-dimensional 
non-interacting electrons into those of a
zero-dimensional system.  However,
interactions couple the orbital states within a Landau level 
and, as we shall see, the similarity to zero-dimensional systems is lost 
when interactions are important.
We can invert Eq.\ (\ref{eq:fillni}) to express $z$ in terms of 
$\nu$.  Substituting the result in Eq.\ (\ref{eq:omegani}) yields
$\Omega_{NI} =  N_{\phi} k_B T \ln (1 - \nu )$ which can be
expanded to identify the virial coefficients in the non-interacting
case:
\begin{equation}
a_l^{NI} = 1/l.
\label{eq:alni}
\end{equation}
Similar elementary calculations yield $g_T^{NI} = N_{\phi} \nu (1 - \nu )/k_B T$
and $F_{NI} = -T S_{NI} =  N_{\phi} k_B T [\nu \ln \nu 
+ (1 - \nu) \ln (1 - \nu) ] $.  It follows from these relations that  
$g_1^{NI}=1$ , $g_2^{NI}=-1 $ and $g_l^{NI} = 0
$ for $l \ge 3$ and that 
$f_1^{NI} = -1 $ and $f_l^{NI} = 1/l(l-1)$ for $l \ge 2$.
Note that the heat capacity vanishes and that the free energy 
is purely entropic in this model since the internal energy is zero
at all temperatures.
 
In closing this section we remark on the particle-hole symmetry 
which exists in the lowest Landau level.  It follows from
this symmetry that\cite{leshouches} the chemical potentials
at filling factors $\nu$ and $1 - \nu$ are related by 
$\mu(\nu) + \mu(1 - \nu) = 2 \epsilon_0$ where $\epsilon_0$ 
is the energy per-electron in the full Landau level state.
$\epsilon_0$ is readily evaluated for any model electron-electron
interaction; for the hard-core model discussed later in this 
article $\epsilon_0 = 4 V_1$.  When expressed in terms of the 
fugacity this relation takes the form $ z(\nu)  z ( 1 -\nu)
= \exp ( 2 \epsilon_0/ k_B T) $.  The virial 
expansions discussed in this paper converge more rapidly at 
lower filling factors and higher temperatures.  This particle-hole
symmetry relation allow us to restrict our attention to the 
regime where $\nu < 1/2$.  

\section{Virial Expansion for the Infinite Hard Core Model} 

A convenient parameterization\cite{haldane}
of the interaction potential between
a pair of electrons in the lowest Landau level results from the
fact that there is only one possible relative motion state
for each angular momentum.\cite{leshouches}  Since isotropic 
interactions will be diagonal in relative angular momentum, the 
interaction Hamiltonian is completely specified
by the interaction energy eigenvalue, 
$V_l$ associated with each relative angular momentum.  (Only nonnegative 
relative angular momenta occur.)  The $V_l$ parameters are known
as Haldane pseudopotentials.  For spin-polarized states 
only odd relative angular momentum states are permitted by
the antisymmetry requirement on the many-fermion wave function.
In the hard-core model (HCM) a pair of electrons interact with each other
only when they are in the smallest possible relative angular momentum 
state, $l =1$, and the Hamiltonian is proportional to $V_1$.
The Hamiltonian $H$ in this model can be written as 
\begin{equation}
H= V_1 \sum_{i<j}^{} P_{ij}^{(1)}
\end{equation}
Here $P_{ij}^{(1)}$ projects particles $i$ and $j$ onto a state of relative 
angular momentum $1$.  The model is non-trivial since the projection
operators for pairs with a common member do not commute.  
In this section we discuss the thermodynamics of the infinite hard-core
model (IHCM) where $V_1 \to \infty$.  Since $V_1$ is the only energy scale 
in the hard-core model Hamiltonian, $V_1$ can be compared only
with the thermal energy $k_B T$ and the results obtained for
the virial coefficients here are $T \to 0$ limits of the general
results for the hard-core model.

For $V_1 \to \infty$ the only finite energy eigenstates of 
the hard-core model are the zero-energy eigenstates.  The calculations in 
this section require only known\cite{MacDonald}
results for the 
degeneracy of the zero-energy eigenvalues as a function of $N_{\phi}$ and $N$.
We briefly summarize one argument leading to this expression.
For $N_{\phi}$ states in a Landau level, the symmetric gauge  
single-particle eigenstates are $\propto \xi^m \exp (- |\xi|^2/4 )$ 
where we have adopted $ \ell \equiv (A/ 2 \pi N_{\phi})^{1/2}$ as 
the unit of length and $m = 0, \ldots, N_{\phi}-1$.   
Here $\xi = x + i y$  is a complex coordinate.
Zero energy many-body eigenstates of the HCM must have a symmetric-gauge 
many-body wave functions of the form\cite {girvin,leshouches}
\begin{equation}
\Psi [\xi ]=\left(\prod_{k=1}^N\exp (-|\xi_k|^2/4)\right)
U[\xi ]\left(\prod_{i<j}^{N}(\xi_i-\xi_j)^3\right).
\end{equation}
in order to avoid pairs in states of relative angular momentum one.
$U[\xi ]$ is any symmetric polynomial in each $\xi_i$ and can 
be regarded as a many-boson wave function. 
The maximum power to which each $\xi_i$ is raised 
in $\prod_{i<j}^{N}(\xi_i-\xi_j)^3$ is $3(N-1)$ . Since the maximum 
possible power of a $\xi_i$ in $\Psi [\xi ]$ is $N_{\phi}-1$ it
follows that the number of zero energy eigenstates of the HCM,
$g(N_{\phi},N)$ 
is equal to the number of $N$-boson states in which the bosons
are allowed to occupy single-particle states with 
$0 \le m \le N_{\phi}-1 - 3 (N-1)$:  
\begin{equation}
g(N_{\phi},N) ={{N_\phi -2(N-1)}\choose N}.
\label{eq:degeneracies}
\end{equation}
$g(N_{\phi},N)$ can also\cite{MacDonald} be regarded as the 
number of $N$ fermion states in which the fermions are allowed to
occupy single-particle states with $0 \le m \le N_{\phi}-1-2(N-1)$.

We now use Eq.\ (\ref{eq:degeneracies}) to determine the virial
expansion coefficients of the IHCM.  We obtain the expansion coefficients
using two different approaches.  First we obtain the coefficients
using an iterative approach in which the cluster expansion 
coefficients are calculated from partition functions for systems 
containing different numbers of electrons.  Our motivation
here is to illustrate the approach used in the following 
section to treat the HCM at finite temperatures.
The general relationship between the grand partition function ($Z$) 
and the canonical ensemble 
partition function for fixed particle number $N$ ($Q_N$) is
\begin{equation}
Z=\sum_{N=0}^{N_{\rm max}} z^N Q_N
\label{eq:zgdef}
\end{equation}
For the infinite hard-core model the maximum $N$ for states
in the Hilbert space is given by 
$N_{\rm max} = \left[ (N_{\phi}+2)/3 \right]$ and all states have zero 
energy so that $Q_N = g(N_{\phi},N)$.  We define finite-size
cluster coefficients $\tilde b_l$ by requiring Eq.\ (\ref{eq:filling})
to hold for each finite value of $N_{\phi}$.  Then 
\begin{equation}
\langle N \rangle = \frac {\sum_{N=0}^{N_{\rm max}} N Q_N z^N}
{Z} =  N_{\phi}  \sum_{N=0}^{\infty} 
N {\tilde  b}_N z^{N}.
\label{eq:Nfinite}
\end{equation}
Multiplying both sides of Eq.\ (\ref{eq:Nfinite}) by $Z$ and equating
coefficients of $z^N$ we find that 
\begin{equation}
\tilde b_N = \frac{Q_N}{N_{\phi}}
- \sum_{l=0}^{N-1} \frac{l}{N} {\tilde b}_l Q_{N-l}.
\label{eq:biterative}
\end{equation}
Using Eq.\ (\ref{eq:biterative}), knowledge of $Q_l$ for $l = 1,
\cdots, N$ is sufficient to evaluate $\tilde b_l$ for $l = 1,
\cdots, N$.  Extrapolating to $N_{\phi} \to \infty$ gives the 
cluster coefficients $b_l$.  For the IHCM $Q_l = g(N_{\phi},l)$ so
this procedure is readily carried out.  It is easy to 
show that the finite-size corrections to $\tilde b_l$ are 
$\propto N_{\phi}^{-1}$, facilitating the extrapolation to $N_{\phi}
\to \infty$.  The leading cluster coefficients obtained by this 
iterative procedure are listed in Table II.

Thermodynamic properties of the IHCM can be calculated analytically
using several different approaches, one of which we outline below.
For a finite $N_{\phi}$ the filling factor is related to the 
grand partition function by 
\begin{equation}
\nu (N_{\phi}, z)=\frac{z}{N_{\phi}}\frac{dZ/dz}{Z}.
\label{eq:nulogderiv}
\end{equation}
Using the explicit expression for $Q_N = g(N_{\phi},N)$ it
follows that for $N_{\phi} \to \infty$
\begin{equation}
\frac{d \nu}{ d (\ln z)} = 6 \nu^3 - 5 \nu^2 + \nu 
\label{eq:ihcmtdos}
\end{equation}
and hence that 
\begin{equation}
z = \frac{\nu (1 - 2 \nu )^2}{ (1 - 3 \nu)^3}.
\label{eq:zihcm}
\end{equation}   
Eq.\ (\ref{eq:zihcm}) can be solved to express $\nu$ in terms 
of $z$:
\begin{eqnarray}
 \nu(z) & =& \frac{1}{3}-\frac{1}{54^{1/3}(4+27z)} \nonumber \\
& &\ \ \times\left ({\root 3 \of {(4+27z)^2-(27z(4+27z)^3)^{1/2}}}
+{\root 3 \of {(4+27z)^2+(27z(4+27z)^3)^{1/2}}}\right ) 
\label{eq:nuzihcm}
\end{eqnarray}
The cluster expansion coefficients listed in Table II can be 
confirmed by expanding Eq.\ (\ref{eq:nuzihcm}).

The virial expansion coefficients we have defined can 
be calculated from Eq.\ (\ref{eq:zihcm}).  
The chemical potential as a function of filling
factor is given by the equation:
\begin{equation}
\mu(\nu)=k_B T\ln\left (\frac{\nu(1-2\nu)^2}{(1-3\nu)^3}\right )
\label{eq:muihcm}
\end{equation}
This equation shows explicitly that $\mu(\nu)$ diverges as $\nu 
\to 1/3$ in the IHCM.  (For a finite $V_1$ HCM, $\mu$ has a finite jump
discontinuity\cite{gros} at $\nu =1/3$.)  Differentiating
Eq.\ (\ref{eq:muihcm}) with respect to $\nu$ we find an analytic
expression for the thermodynamic density-of-states:
\begin{equation}
g_T = \frac{N_{\phi}}{k_B T} \nu (1  - 2 \nu) ( 1 - 3 \nu).
\label{eq:gtihcm}
\end{equation} 
 From Eq.\ (\ref{eq:gtihcm}) we see that for the 
IHCM $g_1^{IHC}=1$, $g_2^{IHC}=-5$, 
$g_3^{IHC}=6$ and 
$g_l^{IHC} = 0$ for $l \ge 4$.  It is remarkable that this expansion
truncates at a finite order.  The grand potential
follows from Eq.\ (\ref{eq:gtihcm}) by using that 
$N  = -\partial \Omega / \partial \mu $ and integrating over filling
factors:
\begin{equation}
\Omega = - \int_{0}^{N} \frac{ N'}{g_T(N')} d N' 
 = - k_B T N_{\phi} \ln  \left (\frac{1-2\nu}{1-3\nu} \right ).
\label{eq:omihcm}
\end{equation}
 From Eq.\ (\ref{eq:omihcm}) we see that for the IHCM
$a_l^{IHC} = (3^l-2^l)/l$.  This virial expansion converges only
for $\nu < 1/3$ as expected.  Adding $\mu N$ to $\Omega$ we find that 
\begin{equation}
F = - T S = N_{\phi} k_B T \big[\nu\ln\nu - (1-2\nu)\ln(1-2\nu) +
(1-3\nu)\ln(1-3\nu) \big].
\label{eq:fihcm}
\end{equation}
so that $f_1^{IHC}=-1$ and $f_l^{IHC} = (3^l - 2^l) / l (l-1)$ for $l \ge 2$.
As for the non-interacting electron model the free energy is purely
entropic because all states in the Hilbert space have zero energy.  The 
expression for the entropy in Eq.\ (\ref{eq:fihcm}) can
be obtained from large number approximations for $g(N_{\phi},N)$ 
and is the starting point for an alternate canonical ensemble 
derivation of the above expressions.

It is worth commenting on the relationship between the 
statistical mechanics of the IHCM and the statistical mechanics
of exclusion statistics gases.\cite{exclusionstats}
The exclusion statistics concept was first proposed by
Haldane, motivated in part by interest in the statistics of 
the fractionally charged quasiparticles\cite{laughlin} of the fractional 
quantum Hall effect.\cite{halperinstats} In a system with
exclusion statistics parameter $g$ the number of states available to 
an added electron, appropriately defined, is decreased by 
$g$ units for each added particle.  With this definition,
it is readily shown that for the hard-core model
the fractionally charged quasiholes\cite{johnson} of the 
$\nu = 1/3$ state in the fractional quantum Hall effect,
present in low-energy states for $\nu < 1/3$,  
satisfy exclusion statistics with statistics parameter $g=1/3$.
On the other hand Eq.\ (\ref{eq:zihcm}) can be recognized as 
the equation satisfied by the occupation number
distribution function\cite{yswu,nayak,chan,rajagopal,murthy} of
a fractional statistics gas 
with statistics parameter $g=3$.  The difference between these 
statistics parameters is readily understood from the equivalence
between the number of zero-energy states in the IHCM and the 
number of many-boson states in a Hilbert space with 
$N_{\phi} - 3(N-1)-1$ single-particle levels.  Eq.\ (\ref{eq:zihcm}) 
is identical to the corresponding relation for a $g=3$ 
exclusion statistics gas because $3$ single-particle 
states are lost for each added electron.  On the other hand,
when the zero-energy states of the hard-core model are 
described\cite{MacDonald} in the language of fractionally
charged quasiholes, the number of particles is 
$N_{qh} = N_{\phi} - 3 (N-1) - 1$ and the number of 
available states is $N$. 
Quasiholes are added at fixed $N_{\phi}$ by decreasing 
$N$ so that the number of states is decreased by 
one for each three added particles giving exclusion statistics
parameter $g=1/3$.  The fact that IHCM 
is equivalent to both a $g=1/3$ exclusion statistics 
gas of quasiholes and a $g=3$ exclusion statistics gas of 
electrons is related to the exclusion statistics particle-hole 
dualities noted by Nayak and Wilczek\cite{nayak}.
At higher temperatures the description in terms of quasiholes
is no longer useful, hence for our purposes it is more useful to regard the 
zero-temperature limit of the HCM as a $g=3$ exclusion
statistics electron system.  

The results obtained here for the IHCM in which 
$V_1 \to \infty$ can be generalized to models where infinite 
repulsion occurs in all relative motion states 
with odd relative angular momentum 
less than or equal to $m$.  In this case the grand partition function is
\begin{equation}
 Z^{(m)}=\sum_{k=0}^{N_{\rm max}^{(m)}}{{mN_{\rm max}^{(m)}-(m-1)k}\choose k} z^k
\end{equation}
where $N_{\rm max}^{(m)}\equiv [(N_{\phi} +m-1)/m]$.
These models correspond generally to $g=m$ fractional statistics gases
and the results described above have simple correspondences.  
For example, 
\begin{equation}
z=\frac{\nu (1-(m-1)\nu)^{m-1}}{(1-m\nu)^m},
\end{equation}
\begin{equation}
\mu =k_BT\ln\left (\frac{\nu (1-(m-1)\nu )^{m-1}}{(1-m\nu )^m}\right ), 
\end{equation}
and 
\begin{equation}
\Omega^{(m)}=-k_BTN_{\phi}\ln\left (\frac{1-(m-1)\nu}{1-m\nu}\right ).
\end{equation}
The chemical potential now diverges at $T=0$ 
for $\nu\to\frac{1}{m}$.

\section{Numerical results at finite T}

\begin{figure}[b]
\epsfxsize=3.9in
\centerline{\epsffile{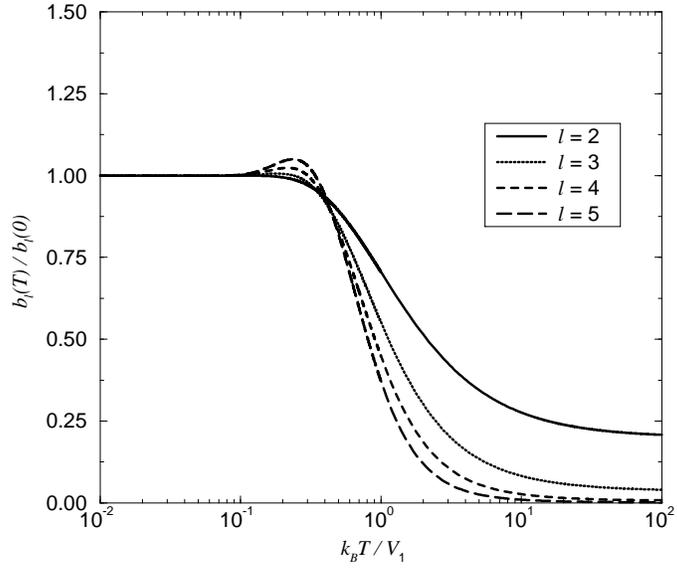}}
\caption{The temperature dependence of the first five cluster coefficients.
	 ($b_1\equiv 1$ and it is not shown.)}
\label{bcoef}
\end{figure}

\begin{figure}[b]
\epsfxsize=3.9in
\centerline{\epsffile{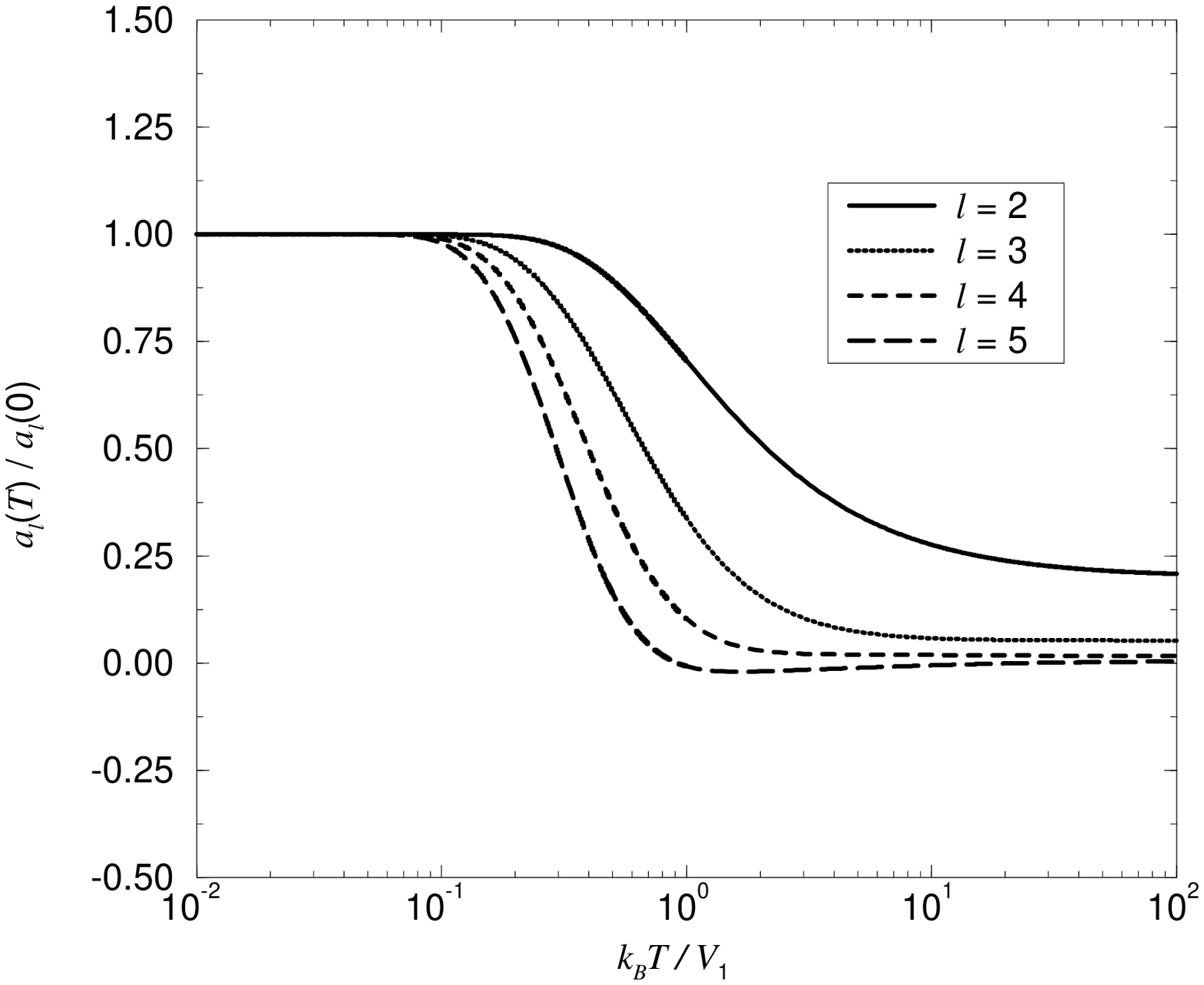}}
\caption{The temperature dependence of the first five virial coefficients.
	 ($a_1\equiv 1$ and it is not shown.)}
\label{acoef}
\end{figure}

\begin{figure}[b]
\epsfxsize=3.9in
\centerline{\epsffile{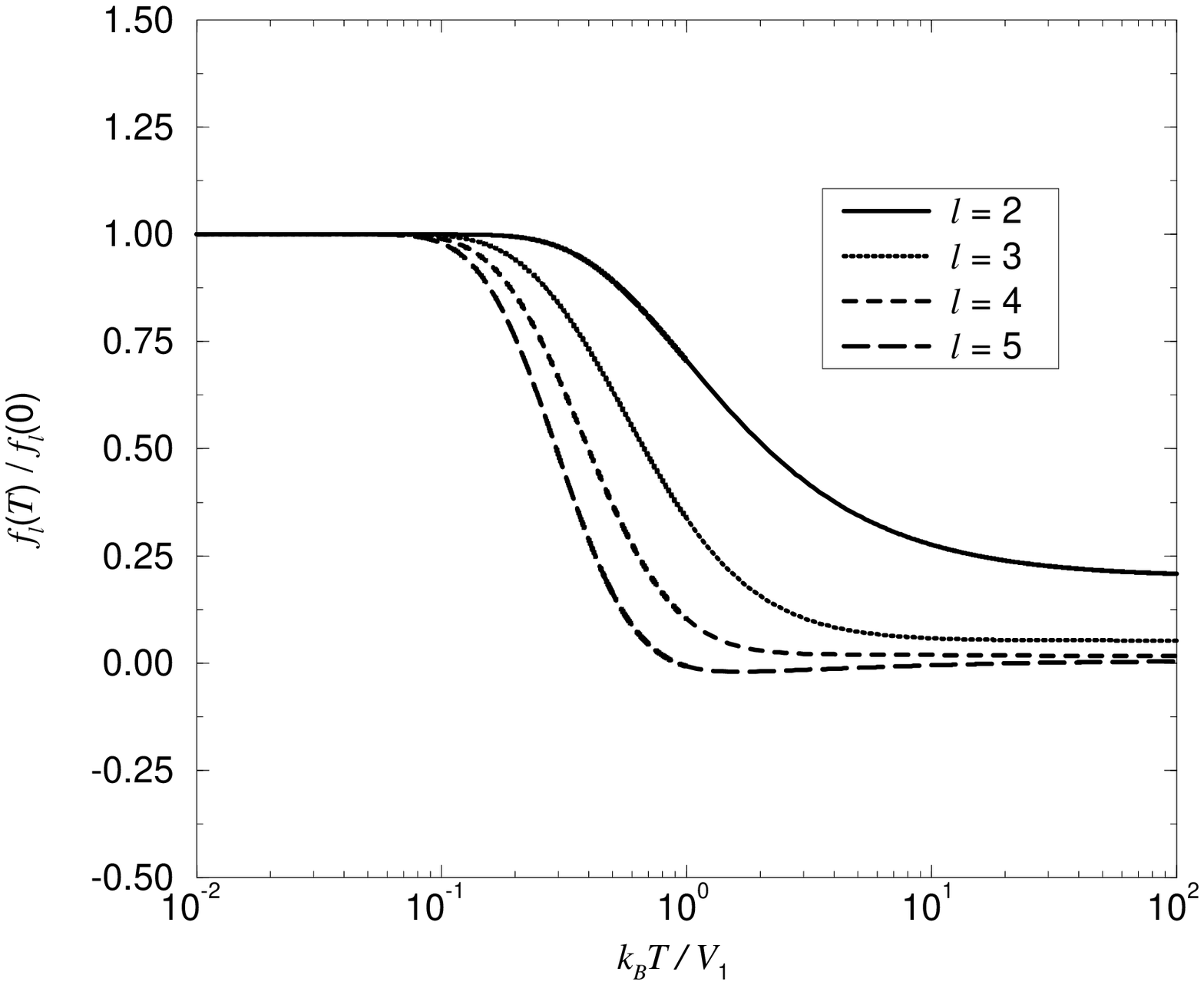}}
\caption{The temperature dependence of the first five $f$-coefficients.
	 ($f_1\equiv -1$ and it is not shown.)}
\label{fcoef}
\end{figure}

\begin{figure}[b]
\epsfxsize=3.9in
\centerline{\epsffile{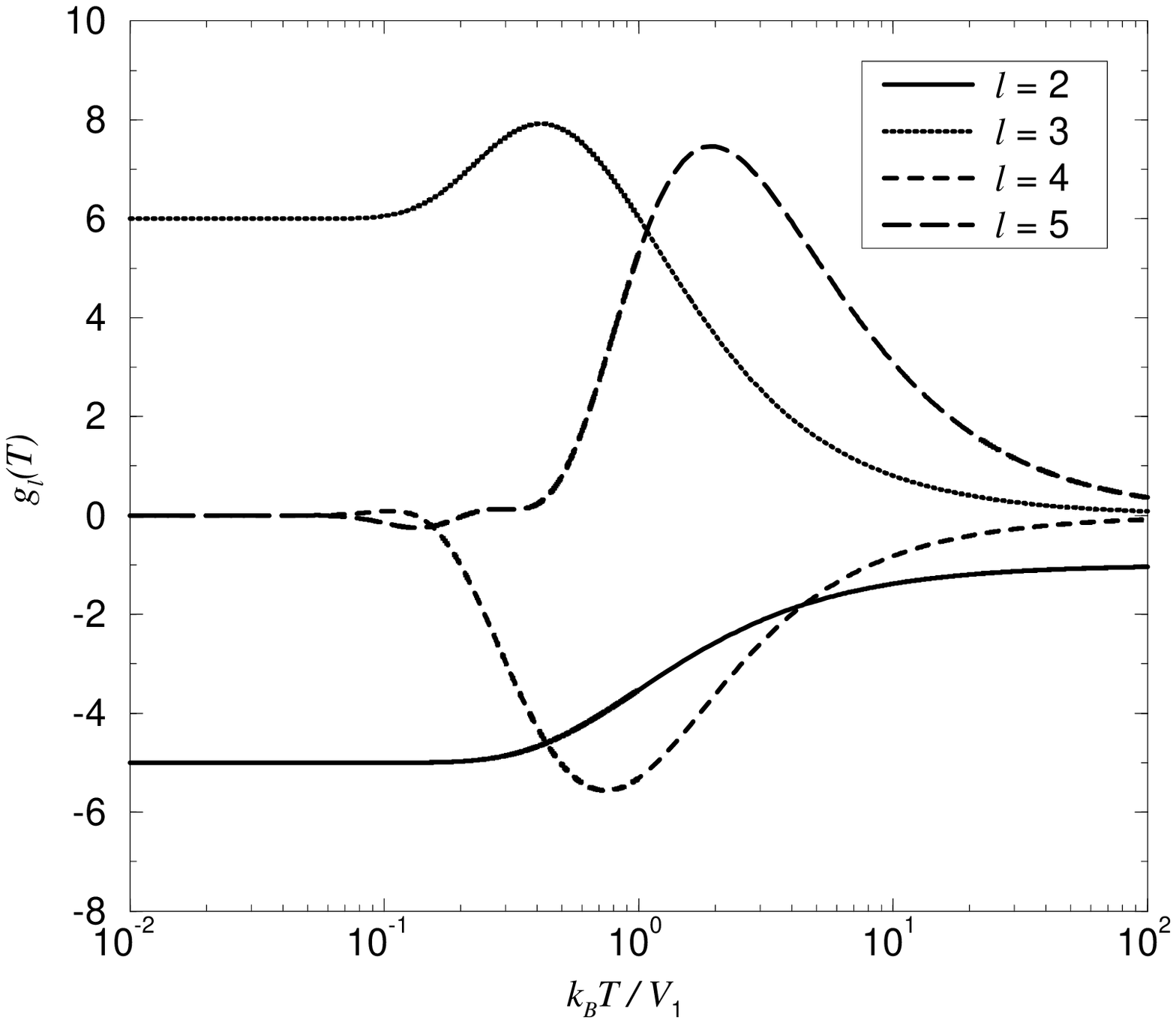}}
\caption{The temperature dependence of the first five $g$-coefficients.
	 ($g_1\equiv 1$ and it is not shown.)}
\label{gcoef}
\end{figure}

In this section we evaluate the expansion coefficients at
finite temperatures assuming HCM electron-electron interactions.
The first two cluster coefficients 
can be calculated analytically giving $b_1(T)=1$ and 
$b_2(T)=-2.5+2\exp (-\beta V_1)$ where $\beta\equiv 1/k_BT$. 
This result for $b_2(T)$ follows from analytic results for 
the complete many-particle energy spectrum and the canonical ensemble
partition function for a two electron problem in a degenerate Landau level. 
To proceed further we numerically
diagonalize the HCM Hamiltonian for finite systems with up to
five electrons and three values of $N_{\phi}=17, 19, 21$. 
The first five cluster coefficients are then evaluated iteratively 
by using Eq.\ (\ref{eq:biterative}) and extrapolating to the limit 
$N_{\phi} \to\infty$.
The leading finite size corrections 
to the virial coefficients are proportional to $N_{\phi}^{-1}$,
as for the infinite hard-core model and the extrapolation
to $N_{\phi} \to \infty$ can be carried out with some 
confidence.  The other expansion coefficients can be obtained
using the relations given in Table I. 
The results are summarized in 
Figs.\ \ref{bcoef} - \ref{gcoef}.  We note that the crossover between low and 
high temperature limits of the cluster coefficients begins for 
$ k_B T \sim 0.1 V_1$.  The chemical potential discontinuity at $\nu =1/3$ in the 
HCM is\cite{gros} $\sim V_1$; it follows that deviations from exclusion
model thermodynamics in the fractional quantum Hall effect start to become important 
at temperatures well below the gap temperature.  We also note that although 
the virial coefficients, $a_l(T)$, are positive for $T=0$ ($a_l = (3^l-2^l)/l$) 
and $T \to \infty$ ($a_l = 1/l$) limits, they do become negative at intermediate
temperatures.  In addition, the higher $l$ coefficients in the virial
expansion of the thermodynamic density of states, which vanish in 
both high temperature and low temperature limits, are comparable to the 
$l=2$ and $l=3$ coefficients at intermediate temperatures.  

\begin{figure}[b]
\epsfxsize=3.9in
\centerline{\epsffile{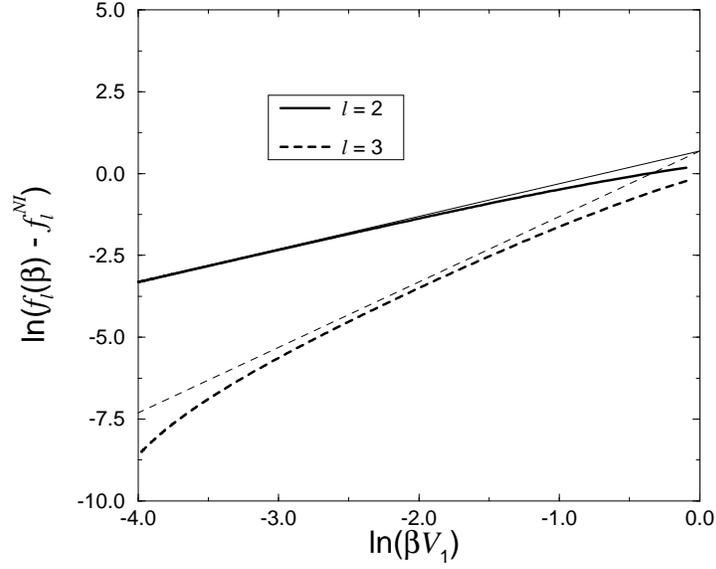}}
\caption{Leading order high-temperature corrections to the non-interacting 
	 Fermi gas values for $f_2(T)$ and $f_3(T)$. The thin lines 
	 correspond to the analytical leading order corrections.}
\label{tdepen}
\end{figure}

We can compare some of these results with 
analytic many-body perturbation theory results\cite{zheng} for 
leading terms in the expansion of the grand potential in powers of $\beta V_1$. 
It follows from that approach that at high temperatures 
\begin{eqnarray}
f_2(T) &=& f_2^{NI} + 2\beta V_1 - (\beta V_1)^2 + (\beta V_1)^3/3 + 
O(\beta V_1)^4,\nonumber\\
f_3(T) &=& f_3^{NI} + 2(\beta V_1)^2 - 140(\beta V_1)^3/81 +
O(\beta V_1)^4,\nonumber\\
f_4(T) &=& f_4^{NI} - (\beta V_1)^2 + 95(\beta V_1)^3/27 +
O(\beta V_1)^4,\nonumber\\
f_5(T) &=& f_5^{NI} - 86(\beta V_1)^3/27 + O(\beta V_1)^4.
\label{eq:largeT}
\end{eqnarray}
Note that the leading correction to $f_2$ is linear in 
$\beta V_1$ while the leading correction to $f_3$ is 
quadratic in $\beta V_1$.   
These corrections can be extracted from our numerical results and they 
are plotted in Fig.~\ref{tdepen}.
For $l=3$ some discrepancy between
the analytic result is visible at the left hand side of this 
log-log plot.  We attribute this discrepancy to errors associated  
with the extrapolation of the numerical results to the thermodynamic limit.
(The discrepancy for $f_3$ at $\beta V_1 = e^{-4}$ is $\sim 0.0005$ 
compared to the $T \to \infty$ value $f_3^{NI} = 1/6$.) 
For $l=4$ and $l=5$ finite size effects are larger so we should 
expect the remnant finite-size error to be larger.  On the 
other hand, the high $T$ corrections to $f_l$ are smaller at a 
given value of $\beta V_1$.  For this reason we have been
unable to establish agreement between our numerical results and 
the coefficients of the leading high temperature correction terms in
these cases.

At low temperatures leading corrections to $f$-coefficients
show activated behavior, with the activation energy decreasing slowly
with increasing $l$.  
In the next section we will attempt to estimate some thermodynamic
properties using information from this virial expansion.  
For $l > 5$ we know only the $T=0$ and $T = \infty$ limits 
of the coefficients; this motivates testing the reliability of simple
formulas for interpolating between the two limits. 
For example, we have examined approximating the temperature dependence of
the $f$-coefficients by the formula
\begin{eqnarray}
f_l(T) = f_l^{IHC} + (f_l^{NI} - f_l^{IHC}) \exp(-\lambda_l\beta V_1)
\label{eq:fit}
\end{eqnarray}
where the $\lambda_l$ are fitting parameters.  
This form is exact for $l=1$ and $l=2$ with $\lambda_1=0$
and $\lambda_2=1$.
The fact that the high temperature corrections to the non-interacting Fermi gas 
values are not captured by these formulas provides a cautionary 
indication that this approach might not be wholly successful.
Using a least squares criterion to fit 
$f_3(T)$, $f_4(T)$, and $f_5(T)$, we find the optimal values 
of $\lambda_3=0.5987$, $\lambda_4=0.5181$, and $\lambda_5=0.4870$.
We note that truncating the virial expansion at fifth order
introduces a relative error of at most 1\% for the free energy
evaluated at $\nu\le 1/6$ and zero temperature.  At higher temperatures 
the virial expansion will converge more rapidly.  Nevertheless, at low temperatures
and filling factors close to $\nu =1/3$ a truncated virial expansion will never
suffice.  In the next section we discuss using interpolation formulas like 
Eq.~(\ref{eq:fit}) to continue the expansion to higher orders.

\section{Thermodynamic Properties From the Virial Expansion}  

\begin{figure}[b]
\epsfxsize=3.9in
\centerline{\epsffile{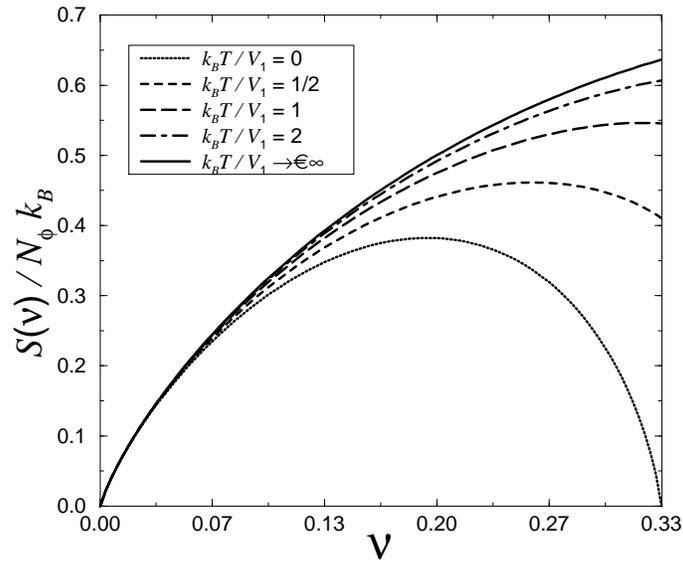}}
\caption{Entropy of a HCM gas vs. filling factor at different
	 temperatures.}
\label{entropy}
\end{figure}

\begin{figure}[b]
\epsfxsize=3.9in
\centerline{\epsffile{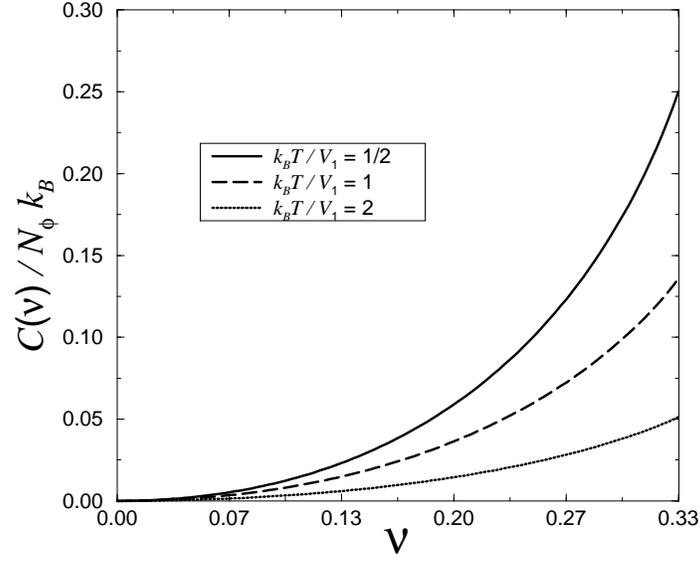}}
\caption{Heat Capacity of a HCM gas vs. filling factor at different
	 temperatures.}
\label{spheat}
\end{figure}

In this section we use information from our virial expansion study to 
estimate some thermodynamic properties in the quantum Hall regime.  
Except where noted, the expansion is carried out to convergence.
At low temperatures this frequently requires virial coefficients 
with indices larger than $l=5$.  In these cases we have approximated 
$f_l(T)$ using Eq.~(\ref{eq:fit}) with $\lambda_l$ extrapolated\cite{note}
from the values obtained by fitting, as described above, to larger 
values of $l$.  
In Fig.~\ref{entropy} the entropy is plotted as a function of filling
factor.  The $T=0$ and $T=\infty$ curves show exclusion statistics
and free particle thermodynamics respectively.  The temperature 
dependence of the entropy is strongest at $\nu =1/3$ where the large
non-interacting electron value is reduced to zero at $T=0$ 
because the Laughlin state\cite{laughlin} is the only thermally
accessible state.  The approximate entropy we obtain using the 
procedures described above has unphysical negative values at 
low temperatures for $\nu > 1/3$.  The naive extrapolation procedure we use 
at large $l$ is not sufficiently subtle to capture the 
chemical potential jump and entropy cusp which occur at $T=0$.
The specific heat is plotted over the same filling factor range in 
Fig.~\ref{spheat}.  Note that the specific heat would be 
strictly zero without electron-electron interactions.  In the fractional
quantum Hall effect regime electrons don't move faster as they heat 
up; instead they are more likely to occupy states in which particle
positions are less strongly correlated and the interaction energy 
increases.

\begin{figure}[b]
\epsfxsize=3.9in
\centerline{\epsffile{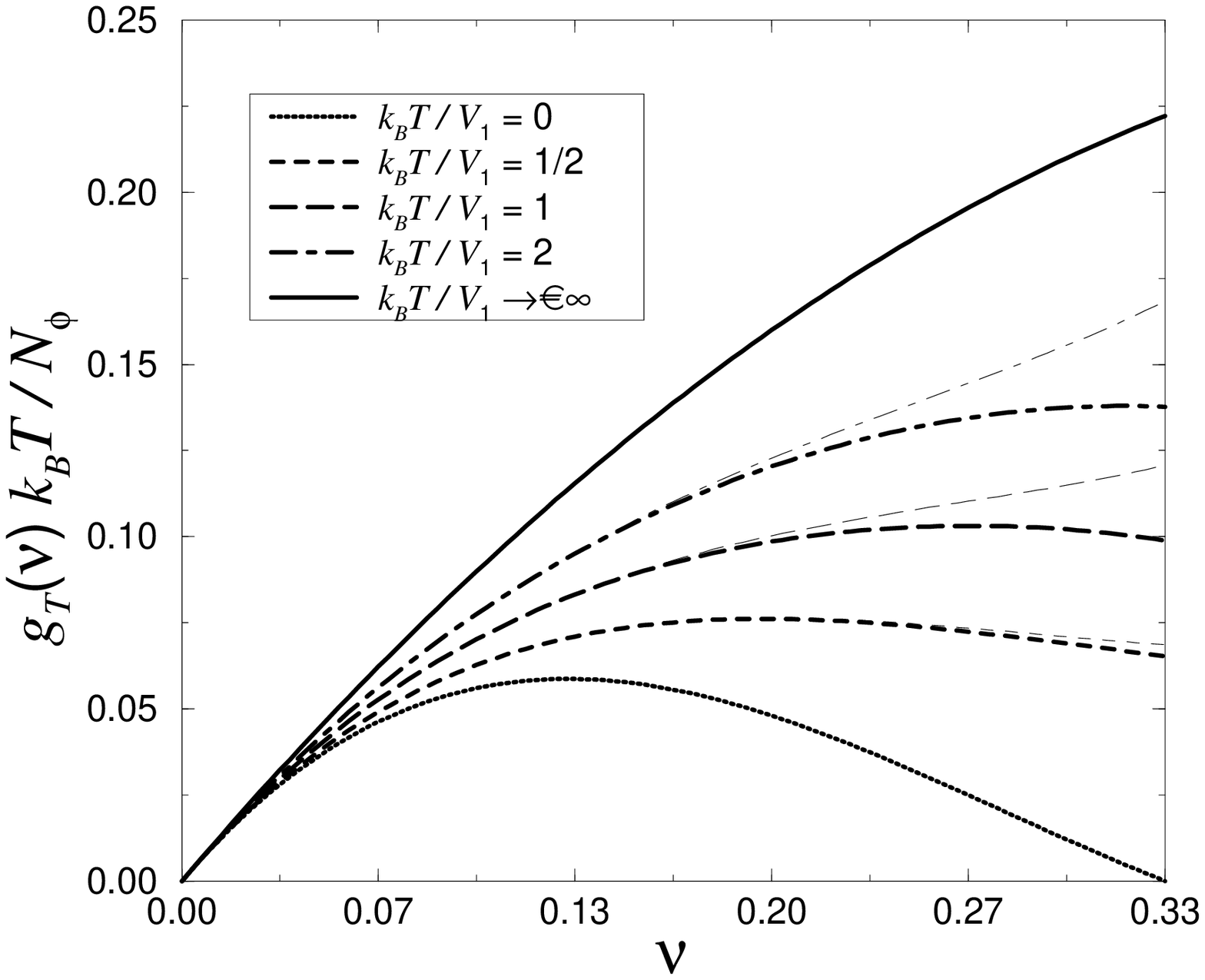}}
\caption{Thermodynamic Density of States of a HCM gas vs. filling 
	 factor at different temperatures. The thick and thin lines 
	 correspond to the expansion truncated at fifth and forth
	 orders, respectively.}
\label{thermodens}
\end{figure}

The thermodynamic density of states is plotted in
Fig.~\ref{thermodens}.  This is one of the most experimentally
accessible thermodynamic properties of two-dimensional electron
systems, thanks to clever capacitative coupling schemes\cite{jimcompress} which take
advantage of the possibility of making separate contact to nearby
two-dimensional electron layers.  In this case the expansion truncates at a 
finite order in both $T=0$ and $T \to \infty$ limits and better
convergence properties allow us to use only the five explicitly
calculated expansion coefficients for the calculation. 
The accuracy of this truncated expansion is indicated by comparing 
with results obtained when only four terms are used in the 
expansion.  The vanishing thermodynamic density of states at zero 
temperature for $\nu=1/3$ is the incompressibility property which is 
central to the fractional quantum Hall effect.  
We find that this basic thermodynamic signature of the fractional
quantum Hall effect is completely absent when $T$ reaches 
half of the gap temperature.  

\section{Summary}

We have examined the virial expansion for a 2D electron gas in the 
fractional quantum Hall regime using a hard-core model for 
electron-electron interactions.  The expansion coefficients 
depend only on the ratio of the hard-core interaction strength $V_1$ 
to the thermal energy $k_B T$.  They may 
be evaluated analytically in both the non-interacting 
high-temperature limit and in the zero-temperature limit where
the thermodynamics is equivalent to that of an 
exclusion statistics gas.  We have evaluated the five leading 
expansion coefficients at intermediate temperatures 
numerically by using an exact diagonalization approach to evaluate the 
canonical ensemble partition function for up to N=5 particles
in finite systems and extrapolating to the thermodynamic limit.
We have used our results to estimate finite temperature properties
of fractional quantum Hall systems and find that the characteristic behavior
of the fractional quantum Hall effect disappears at temperatures 
well below the fractional quantum Hall gap temperature.

This work was supported by the National Science Foundation under grant
DMR-9416906.  The authors are grateful to Lian Zheng for advice on his
high-temperature expansion results.

\begin{table}
\caption{Relations between the first five cluster coefficients and the other 
	 expansion parameters.}
\begin{tabular}{|c|c|c|c|}
$l$ & $a_l$                     & $f_l$                      & $g_l$ \\ \hline

1 & $b_1$                       & $-b_1$                     & $b_1$ \\ \hline
2 & $-b_2$                      & $-b_2$                     & $2b_2$ \\ \hline
3 & $-2b_3+4b_2^2$              & $-b_3+2b_2^2$              & $6b_3-8b_2^2$ \\ \hline     
4 & $-3b_4+18b_2b_3-20b_2^3$    & $-b_4+6b_2b_3-20b_2^3/3$   & $12b_4-48b_2b_3+40b_2^3$ \\ \hline
5 & $-4b_5-144b_2^2b_3$         & $-b_5-36b_2^2b_3$          & $20b_5+360b_2^2b_3$ \\ 
  & $+32b_2b_4+18b_3^2+112b_2^4$& $+8b_2b_4+9b_3^2/2+28b_2^4$& $-112b_2b_4-54b_3^2-224b_2^4$ \\
\end{tabular}
\end{table}

\begin{table}
\caption{Numerical values of the first five expansion coefficients in the IHCM limit.}
     \begin{tabular}{|c|c|c|c|c|}

     $l$ & $b_l^{IHC}$    & $a_l^{IHC}$   & $f_l^{IHC}$        & $g_l^{IHC}$ \\ \hline

      1  & 1              & 1             & $-1$               & 1 \\
      2  & $-5/2$         & 5/2           & 5/2                & $-5$ \\
      3  & 28/3           & 19/3          & 19/6               & 6 \\
      4  & $-165/4$       & 65/4          & 65/12              & 0 \\
      5  & 1001/5         & 211/5         & 211/20             & 0 \\ 
      \end{tabular}
\end{table}

\begin{thebibliography}{99}
\bibitem{note} The extrapolation formula we use is 
$\lambda_l = \sum_{i=1}^4 C_i l^{-i}$ with coefficients $C_i$ obtained using
the least squares approximation for the known values of $\lambda_2$,
$\lambda_3$, $\lambda_4$, and $\lambda_5$: $C_1=-6.013$, $C_2=28.31$, 
$C_3=-59.81$, $C_4=38.44$.
\end{thebibliography}
\end{document}